\documentclass{egpubl}
\usepackage{pg2024}
\usepackage{siunitx}

\SpecialIssuePaper         % uncomment for final version of Computer Graphics Forum, special issue
\CGFStandardLicense
\usepackage[T1]{fontenc}
\usepackage{dfadobe}  

\usepackage{cite}  % comment out for biblatex with backend=biber
\BibtexOrBiblatex
\electronicVersion
\PrintedOrElectronic
\ifpdf \usepackage[pdftex]{graphicx} \pdfcompresslevel=9
\else \usepackage[dvips]{graphicx} \fi

\usepackage{egweblnk} 
\title[FSH3D: 3D Representation via Fibonacci Spherical Harmonics]%
      {FSH3D: 3D Representation via Fibonacci Spherical Harmonics}

\author[Zikuan Li et al.]
{\parbox{\textwidth}{\centering Zikuan Li$^{1}$\thanks{These authors contributed cqually to this work.}\orcid{0000-0001-7471-1188}, Anyi Huang$^{1}$$^\fnsymbol{footnote}$, Wenru Jia$^{1}$, Qiaoyun Wu$^{2}$, Mingqiang Wei$^{1}$, Jun Wang$^{1}$\thanks{J. Wang is the corresponding author.} 
        }
        \\
{\parbox{\textwidth}{\centering $^1$School of Computer Science and Technology, Nanjing University of Aeronautics and Astronautics, China\\
        $^2$ School of Artificial Intelligence, Anhui University, China
       }
}
}
\usepackage{amsmath}
\usepackage{amssymb}
\usepackage{multirow}
\usepackage{color,xcolor}
\begin{document}

% uncomment for using teaser
\teaser{
    \centering
	\includegraphics[width=0.9\linewidth]{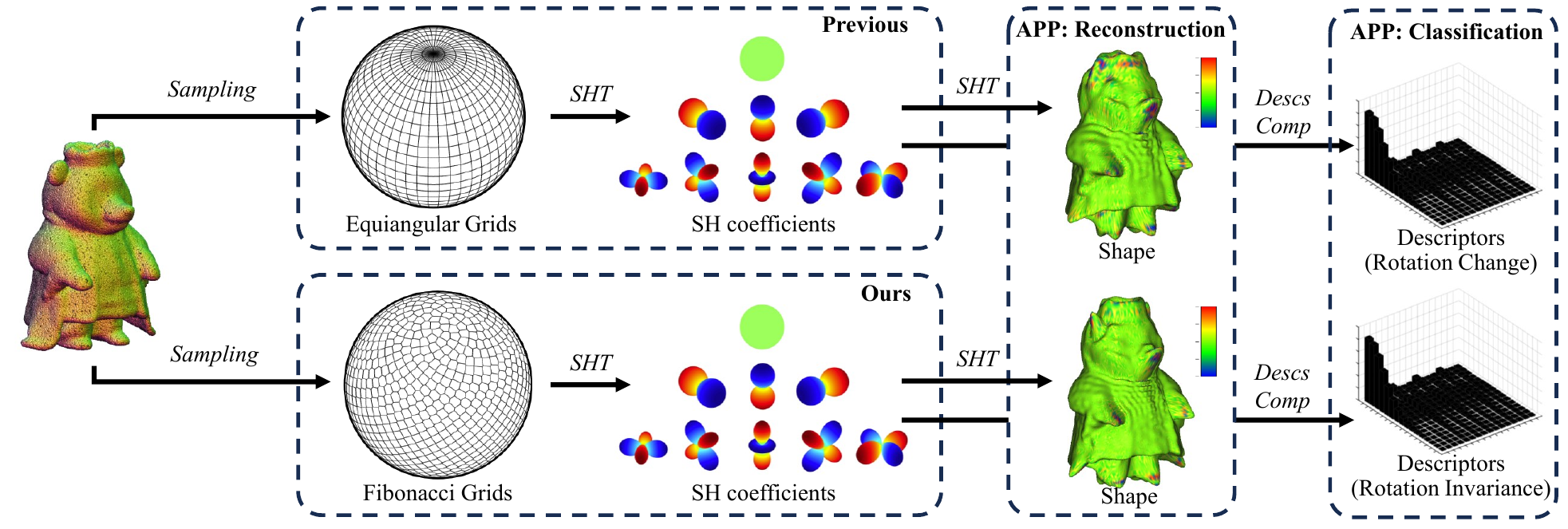}
	\caption{We propose a 3D representation method based on Fibonacci Spherical Harmonics (FSH3D). This method employs a more uniform SFG sampling technique, resulting in a more accurate 3D representation. It enhances the precision of 3D shape reconstruction and improves the rotation invariance of descriptors in 3D shape classification.}
\label{fig:teaser}
}
\maketitle
%-------------------------------------------------------------------------
\begin{abstract}
Spherical harmonics are a favorable technique for 3D representation, employing a frequency-based approach through the spherical harmonic transform (SHT).
Typically, SHT is performed using equiangular sampling grids. 
However, these grids are non-uniform on spherical surfaces and exhibit local anisotropy, a common limitation in existing spherical harmonic decomposition methods.
This paper proposes a 3D representation method using Fibonacci Spherical Harmonics (FSH3D). 
We introduce a spherical Fibonacci grid (SFG), which is more uniform than equiangular grids for SHT in the frequency domain.
Our method employs analytical weights for SHT on SFG, effectively assigning sampling errors to spherical harmonic degrees higher than the recovered band-limited function.
This provides a novel solution for spherical harmonic transformation on non-equiangular grids.
The key advantages of our FSH3D method include: 1) With the same number of sampling points, SFG captures more features without bias compared to equiangular grids; 2) The root mean square error of 32-degree spherical harmonic coefficients is reduced by approximately 34.6\% for SFG compared to equiangular grids; and 3) FSH3D offers more stable frequency domain representations, especially for rotating functions.
FSH3D enhances the stability of frequency domain representations under rotational transformations. 
Its application in 3D shape reconstruction and 3D shape classification results in more accurate and robust representations.
Our code is publicly available at \href{https://github.com/Miraclelzk/Fibonacci-Spherical-Harmonics}{https://github.com/Miraclelzk/Fibonacci-Spherical-Harmonics}.
% \redtext{Our code is publicly available at https://github.com/Miraclelzk/Fibonacci-Spherical-Harmonics.}
%-------------------------------------------------------------------------
%  ACM CCS 1998
%  (see https://www.acm.org/publications/computing-classification-system/1998)
% \begin{classification} % according to https://www.acm.org/publications/computing-classification-system/1998
% \CCScat{Computer Graphics}{I.3.3}{Picture/Image Generation}{Line and curve generation}
% \end{classification}
%-------------------------------------------------------------------------
%  ACM CCS 2012
   % (see https://www.acm.org/publications/class-2012)
%The tool at \url{http://dl.acm.org/ccs.cfm} can be used to generate
% CCS codes.
%Example:
\begin{CCSXML}
<ccs2012>
   <concept>
       <concept_id>10010147.10010371.10010396.10010400</concept_id>
       <concept_desc>Computing methodologies~Point-based models</concept_desc>
       <concept_significance>300</concept_significance>
       </concept>
   <concept>
       <concept_id>10002950.10003714.10003715.10003717</concept_id>
       <concept_desc>Mathematics of computing~Computation of transforms</concept_desc>
       <concept_significance>300</concept_significance>
       </concept>
 </ccs2012>
\end{CCSXML}

\ccsdesc[300]{Computing methodologies~Point-based models}
\ccsdesc[300]{Mathematics of computing~Computation of transforms}

\printccsdesc   
\end{abstract}  
%-------------------------------------------------------------------------
\section{Introduction}
Choosing a 3D representation to describe, model, and understand our world is a crucial topic in computer vision.
The field has recently seen substantial growth due to advancements in deep learning, computational resources, and 3D data availability.
This greatly advances the field of 3D geometry understanding and generation such as retrieval~\cite{nie2023cpg3d}, classification~\cite{ding20233d}, and recognition~\cite{sun2024x}.

3D representations exhibit significant complexity and diversity, necessitating different representations for various geometry processing tasks. 
Historically, explicit representations such as point clouds, voxels, and meshes have been the focus of extensive research~\cite{fan2017point, tan2018variational}. 
These explicit representations are advantageous for rendering and editing purposes due to their straightforward nature. 
However, explicit representations have inherent limitations, including irregularity, ambiguity, and sparsity in representing shapes.
With the advent and rapid progression of deep learning technologies, function-based implicit representations have gained prominence~\cite{park2019deepsdf, mescheder2019occupancy, ouasfi2024unsupervised}. These implicit methods are becoming increasingly popular due to their potential to overcome the limitations of explicit representations.
Spherical harmonic transform (SHT) is a favorable technique for 3D implicit representation that transfers 3D points in the spatial domain to the frequency domain.~\cite{feinauer2015structural}.
It is commonly exploited that many spherical functions in natural phenomena can be sufficiently approximated using only a few spherical harmonic coefficients.
In the computer graphics, spherical harmonics have garnered significant interest due to their contributions in various fields, such as global illumination~\cite{green2003spherical}, shape descriptors~\cite{kazhdan2003rotation}, shape reconstruction of star-shaped point sets~\cite{tosic2006fst}, and 3D representations and filtering~\cite{zhou20043d}.

The SHT is a key step in 3D representations using spherical harmonics.
Driscoll $et$ $al.$ derived a method for calculating the weights of spherical harmonics in the context of equiangular sampling~\cite{driscoll1994computing}. 
This equiangular sampling technique integrates a separation of variables, leading to the refinement of the initial $O(N^4)$ algorithm into a more efficient $O(N \log^2 N)$ algorithm.
SHT methods are generally performed on equiangular sampling. 
However, this sampling is prone to errors related to uneven distribution, with higher density at the poles and sparsity around the equator.
This sampling issue can lead to missing details in the spherical harmonic frequency domain results, particularly when the object rotates. 

To address this problem, one approach is to map the model onto the faces of a suitable polyhedron, with the faces subdivided to form grid cells. 
Grid frameworks derived from a cube~\cite{ranvcic1996global} and those based on a regular icosahedron~\cite{majewski2002operational} are popular. 
However, polyhedral subdivisions have limited ability to precisely control refinement levels. 
Drake $et$ $al.$ introduce a new method for performing spherical harmonic analysis on HEALPix data, with a computational complexity of $O(N \log^2 N)$~\cite{drake2020fast}.
Larkins $et$ $al.$ survey various spherical data structures for normal binning, identifying the spherical Fibonacci grid (SFG) as the most accurate method~\cite{larkins2012analysis}.  
While SFG provides a more uniform scheme, addressing the weight problem is crucial when applied to SHT. 
Swinbank $et$ $al.$~\cite{swinbank2006fibonacci} approximate the weight of each point on SFG as equal weights. 
Ahmad $et$ $al.$~\cite{ahmad2007quasi} define the area of the Voronoi diagram as the weight of each point on SFG, termed area weights. 
However, determining Fibonacci weights under the SHT framework poses unique challenges, requiring energy adjustment considerations for each spherical harmonic degree.

This paper proposes a SHT method based on SFG for more accurate 3D representation, called FSH3D.
We address the Fibonacci sampling weight fixing problem under the SHT framework and provide a weight calculation method. 
This weight allocation effectively redistributes sampling errors to spherical harmonic degrees higher than the recovered band-limited function.
We experimentally validate the stability of the weights in SHT and apply them to 3D shape representation and 3D shape classification.
Our method demonstrates significant improvements over equiangular grids in both applications. 
Our main contributions are summarized as follows:

\begin{itemize}
\item 
We propose a 3D representation using FSH3D.
The SFG can sample more features without bias compared to the equiangular grid when using the same number of sampling points.
FSH3D enhances the stability of frequency domain results, especially when the objective function rotates.
\item 
We present a method for calculating the weights of SFG sampling.
This method allocates sampling errors to degrees higher than the recovered band-limit function, making the Fibonacci SHT more accurate.
\item
The application of FSH3D in reconstructing shapes with star-shaped sets outperforms conventional spherical harmonic computation methods.
The more uniform SFG sampling preserves a higher level of detail in the reconstructed outcomes.
Furthermore, using FSH3D for classification through spherical harmonic descriptors shows superior performance compared to traditional spherical harmonic computation approaches.
\end{itemize}

\section{Related Works}
\subsection{3D Representations}
We describe several of the most commonly used 3D representations.
Point clouds represent 3D shapes using discrete data points in space.
Although these representations~\cite{li2020end,fei2022comprehensive} have a low storage footprint, they lack surface topology information.
Voxel grids~\cite{jiang2020sdfdiff,schwarz2022voxgraf} consist of unit cubes in 3D space, analogous to pixels in 2D images~\cite{xu2023survey}.
Each voxel encodes occupancy information and can also encode the distance to a surface, as seen in the Signed Distance Function (SDF)~\cite{jiang2020sdfdiff} or Truncated Signed Distance Function (TSDF)~\cite{mittal2022autosdf}.
However, voxel grids require substantial memory for high-resolution details.
Polygonal meshes use vertices and surfaces to compactly describe complex 3D shapes.
While they are efficient in terms of storage, they are unstructured and non-differentiable, posing challenges for integration with neural networks in end-to-end differentiable pipelines~\cite{kato2020differentiable}.
While these representations are efficient in terms of storage, they are unstructured and non-differentiable, posing challenges for integration with neural networks in end-to-end differentiable pipelines.
Neural radiance fields (NeRF)~\cite{xie2022neural} have gained significant interest in the 3D research community.
NeRF maps spatial coordinates to scene properties such as occupancy, color, or radiance. 
Unlike traditional representations that rely on geometric primitives, neural fields use a multi-layer perceptron for this mapping~\cite{tewari2020state}.
3D Gaussian splatting extends point clouds by including additional information at each point~\cite{kerbl20233d}. 
This method achieves state-of-the-art novel-view synthesis efficiently by employing rasterization rather than ray tracing, as used in NeRF.

\subsection{Spherical Harmonics Transform}
SHT can effectively approximate various spherical functions in natural phenomena using only a few coefficients. Spherical harmonics are widely used in diverse fields such as Earth's gravitational and magnetic fields, physical chemistry for atomic electronic configuration representation, and computer graphics.
They are employed for precomputing lighting contributions and sampling lighting contributions~\cite{green2003spherical}.
Spherical harmonics have also proven robust in rotation estimation~\cite{althloothi2013robust}. 
Additionally, spherical harmonics are instrumental in constructing 3D shape descriptors and have been successfully used in shape-matching and retrieval scenarios~\cite{papadakis2007efficient, funkhouser2003search}.
Furthermore, Zhou $et$ $al.$~\cite{zhou2015micromorphology} propose a method for micromorphology characterization and reconstruction of sand particles using micro X-ray tomography and spherical harmonics. 
Fabry $et$ $al.$~\cite{fabry2010surface} propose a human face recognition method based on spherical harmonics.
Currently, spherical harmonics are also being combined with deep learning. 
The convolution theorem~\cite{driscoll1994computing} states that spherical convolution is equivalent to point-wise multiplication of harmonic coefficients and can be independent of a neighborhood definition. Spherical convolution can be achieved by transforming either a function on $SO(3)$~\cite{s.2018spherical} or a spherical function~\cite{esteves2018learning} via SHT.

\subsection{Sampling on Sphere}
The challenge of evenly distributing points on a sphere, a longstanding problem, has crucial implications in numerous scientific fields. 
One common method for implementing spherical sampling involves mapping the model onto the faces of a polyhedron, such as mapping to a cube~\cite{ranvcic1996global} or an icosahedron~\cite{majewski2002operational}.
However, polyhedral subdivisions have limited ability to precisely control refinement levels.
Another solution is spherical sampling based on a spiral pattern. 
HEALPix is a versatile structure for the pixelization of data on the sphere~\cite{gorski2005healpix}.
SFG methods have been widely used for near-optimal solutions~\cite{newell2005plants}. 
While SFG provides a more uniform scheme, addressing the weight problem is crucial when applied to SHT.
Swinbank $et$ $al.$~\cite{swinbank2006fibonacci} approximate the weight of each point on SFG as equal weights.
Ahmad $et$ $al.$~\cite{ahmad2007quasi} define the area of the Voronoi diagram as the weight of each point on SFG, termed area weights. 
Reeger $et$ $al.$~\cite{reeger2016numerical} presents a novel approach to create orthogonal weights for N arbitrarily dispersed nodes by using finite differences generated based on radial basis functions.
However, determining Fibonacci weights under the SHT framework poses unique challenges, requiring energy adjustment considerations for each spherical harmonic degree.

\section{Method}
\subsection{Overview}
\begin{figure*}[!t]
    \centering
	\includegraphics[width=1\linewidth]{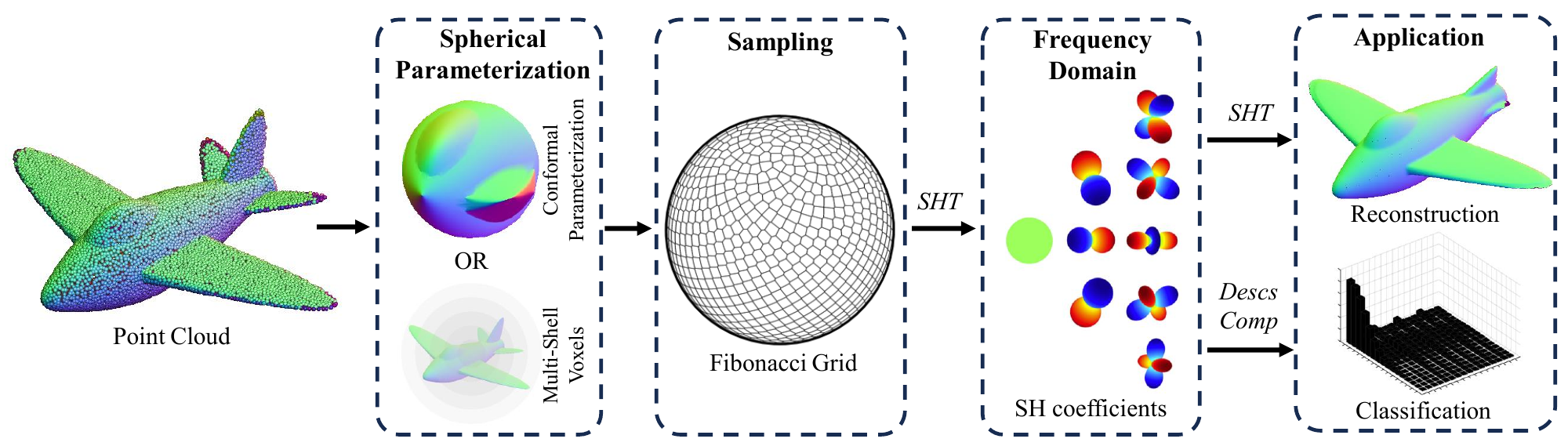}
	\caption{The FSH3D pipeline for 3D representation. Spherical parameterization can be achieved via conformal parameterization or multi-shell voxels, followed by sampling using SFG, and subsequent decomposition over the spherical domain through SHT. Two typical application scenarios are illustrated.
 }
\label{fig:overview}
\end{figure*}
% OverView
To address the challenge of equiangular sampling in the SHT, we propose an alternative approach based on the SFG, called FSH3D, which offers improved accuracy in 3D representation.
The SFG can sample more features without bias than the equiangular grid when using the same number of sampling points.
Fig.~\ref{fig:overview} displays FSH3D framework for 3D representation, encompassing spherical parameterization, sampling, frequency domain representation, and demonstrating two typical applications: reconstruction and classification.
Reconstruction on a unit sphere is facilitated by the SHT~\cite{tosic2006fst}.
Energy descriptors derived from spherical harmonic coefficients are utilized to achieve a rotation-invariant 3D representation~\cite{kazhdan2003rotation}, useful for classification or retrieval tasks.
This article primarily focuses on SFG sampling and the weight solution method for SHT based on SFG. Therefore, detailed discussion on spherical parameterization and its applications is omitted, and readers are referred to relevant literature for further insights~\cite{agus2020wish}.

Here, we first provide a detailed introduction to spherical harmonics, the primary technology used in our paper.
Then we introduce the spherical uniform sampling method using the SFG.
Next, we analyze the discretization problem in the SHT and identify the necessary conditions for implementing SHT with SFG. 
Finally, we present the method for calculating the weights for SHT based on SFG and provide the final formula.

\subsection{Background of spherical harmonics}
The spherical harmonics are eigenfunctions of the Laplace operator on the unit sphere $\mathbb{S}^2$.
Employing a consistent frequency domain representation proves highly advantageous for describing spherical data effectively~\cite{liu2014rotation}. 
Analogous to the Fourier transform, the SHT facilitates the projection of spatial data into the frequency domain. 
Due to this excellent property, 3D signal can be computed to derive corresponding spherical harmonic coefficients, enabling the recovery of specified degree 3D signal.
The spherical harmonic $Y_{l}^{m}(\theta, \phi)$ is defined on the unit sphere:
\begin{equation}
Y_l^m(\theta, \phi)=(-1)^m \sqrt{\frac{2l+1}{4\pi} \frac{(l-m) !}{(l+m) !}} P_l^m(\cos \theta) e^{i m \phi},
\label{Eq1}
\end{equation}
for degree $l \geq 0$ and order $\left| m \right|\le l$. 
The associated Legendre functions on $\left[ -1,1 \right]$ are denoted by $P_{l}^{m}$.
The spherical harmonic bases are consistently indexed by two integers, $l$ and $m$, where $-l\le m\le l$.
Therefore, there are $2l+1$ coefficients for each $l$. 
The frequency of the basis is determined by $l$, while $m$ is just the ``state'' of the spherical harmonic, as expounded in quantum mechanics~\cite{davis2011remarks}.

The expression of a spherical function as a linear combination of spherical harmonics constitutes the SHT. 
Initially, the spherical function $f(\theta, \phi)$ is defined, and its expansion is given by:
\begin{equation}
f(\theta, \phi)=\sum_{l=0}^{\infty} \sum_{m=-l}^{l} \hat{f}_{l}^{m} Y_{l}^{m}(\theta, \phi),
\label{Eq2}
\end{equation}
where $\theta \in[0, \pi],\phi \in[0,2 \pi]$ are the zenith angle and azimuth angle of the spherical coordinate system, respectively. 
Specifically, the spherical harmonic coefficient $\hat{f}_{l}^{m}$ is computed as:
\begin{equation}
\hat{f}_{l}^{m}=\int_{0}^{2 \pi} \int_{0}^{\pi} f(\theta, \phi) \overline{Y_{l}^{m}}(\theta, \phi) \sin \theta d \theta d \phi.
\label{Eq3}
\end{equation}

For a comprehensive understanding of spherical harmonics theory, refer to~\cite{arfken2011mathematical}.

\subsection{Spherical Fibonacci Grids on Sphere}
This work discusses sampling on the space $\mathbb{S}^2$, aiming for uniform discretization across the sphere with equal division into parts.
A uniform spherical distribution is mathematically defined to maximize the minimum distance between points, termed the Tammes problem, a special case of the dense paving problem. 
However, the latter often lacks an elegant solution.

\begin{figure*}[!t]
    \centering
	\includegraphics[width=1\linewidth]{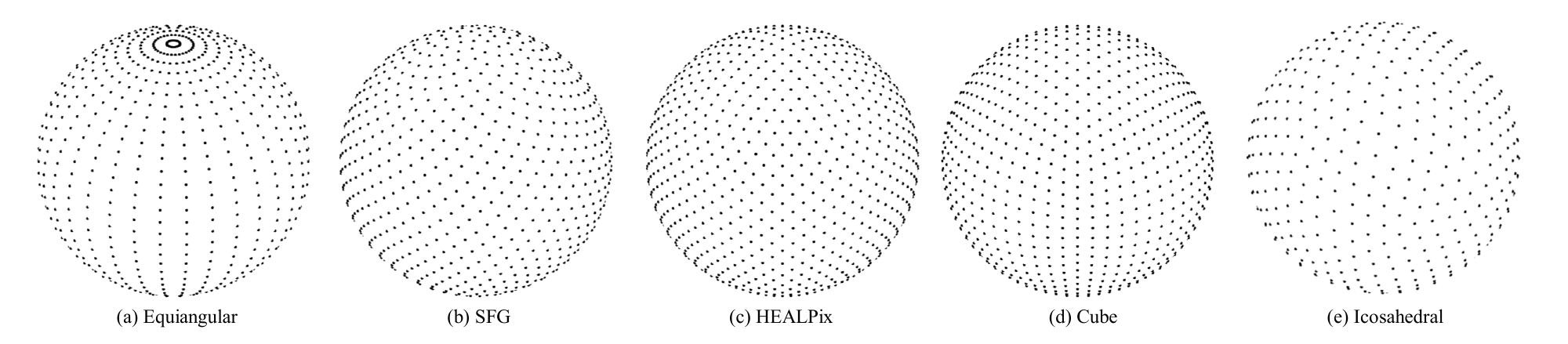}
	\caption{The distribution of different grid methods. (a-e) equiangular grid, SFG, HEALPix grid, Cube gird and Icosahedral grid. }
\label{fig:Distribution}
\end{figure*}

At present, the prevalent sampling method employs equiangular grids, dividing the latitude and longitude directions into equal angles to serve as sampling points. 
However, this method yields non-uniform samples on the sphere, resulting in dense sampling near the poles and sparse sampling near the equator. 
Consequently, with an equal number of sampling points, this approach might overlook certain features, leading to an inability to fully represent the original function. 
To address this issue, researchers have proposed various solutions, among which the SFG stands out for its ease of construction and lesser axial anisotropy. 
Notably, it can uniformly distribute samples in any direction, as depicted in Fig.~\ref{fig:Distribution}, presenting significant advantages over equiangular grids.

The SFG, named after the Fibonacci ratio, is constructed by forming a spiral line around the sphere from the North Pole to the South Pole, dividing the sphere into thin slices of uniform thickness~\cite{swinbank2006fibonacci}. 
Each point is incrementally placed along the spiral line, as depicted in Fig.~\ref{fig:Distribution}(b). 
With a sufficient number of Fibonacci points, nearby points form a crisscross spiral, resulting in an evenly distributed arrangement in all directions, exhibiting uniform, dense, and chaotic characteristics. 
$n$ Fibonacci points are uniformly generated on the sphere, with the SFG distribution $F\left[ \theta ,\phi \right]$ in the spherical coordinate system expressed as follows:
\begin{equation}
\begin{aligned}
{{F}_{\left[ \theta ,\phi  \right]}}\left( i \right)=\left[ {{\sin }^{-1}}\left( \frac{2d}{n} \right)+\frac{\pi }{2},2\pi mod\left( d,\tau  \right) \right],  \\
d=\frac{1-n}{2}+i,i\in 0,1,\ldots ,n-1,
\end{aligned}
\label{Eq4}
\end{equation}
where $mod(d,\tau)$ means taking the decimal part, $\tau=\frac{\sqrt{5}-1}{2}$. 
Utilizing the SFG for measuring irregular shapes' area on the sphere results in a reduction of the root mean square error by at least 40\% compared to weighted equiangular grids~\cite{gonzalez2010measurement}.

\subsection{Discretization Problem of SHT}
The core challenge in realizing the SHT based on the SFG lies in addressing the discretization problem. 
From an information-theoretic viewpoint, the fundamental property of any sampling theorem is the number of samples required to capture all of the information content of a band-limited signal. 
Shannon gives the classical sampling theorem and states that the Fourier transform is a function with bounded support. 
It can represent from its samples, provided that the uniform sample frequency is at least twice the boundary frequency \cite{tain1982optimal}. 
To represent exactly a signal on the sphere band-limited at $b$, all sampling theorems on the sphere require $O ({b}^{2})$ samples. 
The canonical Driscoll \& Healy sampling theorem~\cite{driscoll1994computing} requires $4{{b}^{2}}$ samples at least, and the McEwen \& Wiaux sampling theorem~\cite{mcewen2011novel} requires $2{{b}^{2}}$ samples only. 
For the number of SFG sampling points, we apply the Driscoll \& Healy sampling theorem.

Before solving the discretization problem, it is imperative to elucidate a fundamental theorem concerning spherical harmonic \cite{driscoll1994computing}.

\newtheorem{thm1}{Theorem}
\begin{thm1}\label{thm:t1}
	\begin{equation}
		\begin{aligned}
	Y_{l_1}^{m_1}Y_{l_2}^{m_2}={} \underset{L=\left| {{l}_{1}}-{{l}_{2}} \right|}{\overset{{{l}_{1}}+{{l}_{2}}}{\mathop \sum }}\,\sqrt{\frac{\left( 2{{l}_{1}}+1 \right)\left( 2{{l}_{2}}+1 \right)}{4\pi \left( 2L+1 \right)}}C_{0,0,0}^{{{l}_{1}},{{l}_{2}},{{l}_{3}}}C_{{{m}_{1}},{{m}_{2}},{{m}_{1}}+{{m}_{2}}}^{{{l}_{1}},{{l}_{2}},L}Y_{L}^{{{m}_{1}}+{{m}_{2}}},
	\end{aligned}
	\label{Eq5}
	\end{equation}
	where $C_{{{\mu }_{1}},{{\mu }_{2}},\mu }^{{{\lambda }_{1}},{{\lambda }_{2}},\lambda }$ are Wigner symbols. When $\left| {{m}_{1}}+{{m}_{2}} \right| > L$, the value of $Y_{L}^{{{m}_{1}}+{{m}_{2}}}$ is 0.
\end{thm1}

The SHT is performed under a continuous function, but in practical applications, we can only represent the band-limit function completely with limited sampling. 
We define ${{f}_{s}}$ as a discrete sampling function:
\begin{equation}
{{f}_{s}}=f\cdot s,
\label{Eq6}
\end{equation}
where $f$ is the original function and $s$ is the discrete unit impulse function on the sphere. 
At this time, if $l<b$, ${{\hat{f}}_{s}}\left( l,m \right)=\hat{f}\left( l,m \right)$, then we can consider that the original function $f$ can be represented by ${{f}_{s}}$ which expands not higher than $b$ degree of SHT. 
Therefore, we must construct a reasonable $s$ so that Eq.~(\ref{Eq6}) satisfies this condition.

Discrete samples on the unit sphere, denoted as $s$, undergo the SHT to obtain $\hat{s}$. 
The 0-degree spherical harmonic corresponds to the unit sphere, hence only the zeroth degree of $\hat{s}$ equals 1, with all other degrees being 0.
As the represent process involves a band-limited function, it is essential to ensure that only finite degrees satisfy this condition.
Let the cut-off degree be denoted as $e$, such that $\hat{s}\left( 0,0 \right)=1$ and $\hat{s}\left( l,m \right)=0$ for $0<l<e$.
Thus, $s$ is expressed using spherical harmonics according to these conditions:
\begin{equation}
s=1+\underset{j\ge e}{\mathop \sum }\,\underset{\left| k \right|\le j}{\mathop \sum }\,\hat{s}\left( j,k \right)Y_{j}^{k}\left( \theta ,\phi  \right).
\label{Eq7}
\end{equation}
The determination of the cut-off degree is elucidated as follows:
\begin{align*}
{{f}_{s}} = {} & f+f\underset{j\ge e}{\mathop \sum }\,\underset{\left| k \right|\le j}{\mathop \sum }\,\hat{s}\left( j,k \right)Y_{j}^{k}\left( \theta ,\phi  \right) \\
= {} & f+\underset{j\ge e}{\mathop \sum }\,\underset{\left| k \right|\le j}{\mathop \sum }\,\hat{s}\left( j,k \right)fY_{j}^{k}\left( \theta ,\phi  \right) \\
 = {} & f+\underset{l<b}{\mathop \sum }\,\underset{j\ge e}{\mathop \sum }\,\underset{\left| m \right|\le l}{\mathop \sum }\,\underset{\left| k \right|\le j}{\mathop \sum }\,\hat{s}\left( j,k \right)\hat{f}\left( l,m \right)  Y_{l}^{m}\left( \theta ,\phi  \right)Y_{j}^{k}\left( \theta ,\phi  \right).
\end{align*}

Thus, the sampling function can be decomposed into a band-limited original function and the aliasing error introduced by sampling ${{f}_{s}}$. 
According to Theorem~\ref{thm:t1}, the product of spherical harmonics $Y_{l}^{m}Y_{j}^{k}$ can be expressed as a sum of a series of  $Y_{L}^{m_1+m_2}$, where $\left| l-j \right|\le L\le l+j$. 
To effectively represent the band-limited function, the aliasing error must be confined to spherical harmonics greater than or equal to $b$, implying that $\left| l-j \right|\ge b$.
With the constraints of $j$ and $l$, this condition holds true only when $e\ge 2b$. 
Therefore, we conclude that when at least
\begin{equation}
\hat{s}\left( 0,0 \right)=1,\hat{s}\left( l,m \right)=0,~\text{for}~0<l<2b.
\label{Eq8}
\end{equation}
The band-limited function $f$ with a bandwidth $b$ can be accurately represented by SHT of the sampling function ${{f}_{s}}$.

\subsection{Determine the Weight of SHT}

We employ a more uniform SFG ${{F}_{\left[ \theta ,\phi  \right]}}$ (see Sec 3.3) as the sampling module for the SHT (see Sec 3.2).
Typically, the SFG utilizes equal weights for calculations \cite{swinbank2006fibonacci}.
However, each grid point of the SFG's bounding area contains approximately 10 grid points at the extreme points, with a deviation of around 2\% \cite{swinbank2006fibonacci}.
Consequently, some researchers advocate using area weights for calculations \cite{ahmad2007quasi}. 
Nonetheless, when employing SFG within the SHT framework, it is essential to consider the energy of each degree of spherical harmonics. 
Hence, the weight cannot be directly determined by equal weights or area weights.
Therefore, we utilize weighted compensation to construct the sampling function, ensuring that the sampling meets the conditions specified in Eq.~(\ref{Eq8}).

We define ${ w_{k}^{( b )} }$ as the compensation weight, and let $f\left( \theta ,\phi \right)$ represent a band-limited function with a bandwidth $b$. 
When $l\ge b$, $\hat{f}\left( l,m \right)=0$.
According to the Driscoll \& Healy sampling theorem, for any positive integer $b$, the number of sampling points $n$ must exceed $4b^2$.
We introduce the SFG distribution with a bandwidth of $b$ as:
\begin{equation}
{{s}_{b}}=\sqrt{2\pi }\underset{j=0}{\overset{n-1}{\mathop \sum }}\,w_{j}^{(b)}\delta ( {{F}_{\left[ \theta ,\phi  \right]}}\left( j \right) ),
\label{Eq9}
\end{equation}
where $\delta$ is the Dirac delta function.
We transform the distribution in terms of spherical harmonic coefficient:
\begin{equation}
{{{\hat{s}}}_{b}}\left( l,m \right)=\sqrt{2\pi }\underset{j=0}{\overset{n-1}{\mathop \sum }}\,w_{j}^{(b)}{{\delta }_{\left( {{\theta }_{j}},{{\phi }_{j}} \right)}} ( \overline{Y_{l}^{m}} ).
\label{Eq10}
\end{equation}

Equiangular grids are uniformly distributed in the $\theta $ and $\phi $ directions, facilitating the derivation of a unique analytical solution. 
In contrast, the SFG distribution rule poses challenges in obtaining a closed solution.
Therefore, we employ the least-squares method to determine the weights.

In order to represent the band-limited function using SFG, it is imperative to ensure that the condition in Eq.~(\ref{Eq8}) is met. 
The simultaneous equations are set up to solve the weights according to Eq.~(\ref{Eq10}):
\begin{equation}
\left\{ \begin{matrix}
\sqrt{2\pi }\underset{j=0}{\overset{n-1}{\mathop \sum }}\,w_{j}^{(b)}\delta ( F\left( j \right) ) ( \overline{Y}_{0}^{0} )=1  \\
\sqrt{2\pi }\underset{j=0}{\overset{n-1}{\mathop \sum }}\,w_{j}^{(b)}\delta ( F\left( j \right) ) ( \overline{Y}_{1}^{-1})=0  \\
\begin{matrix}
\vdots   \\
\sqrt{2\pi }\underset{j=0}{\overset{n-1}{\mathop \sum }}\,w_{j}^{(b)}\delta ( F\left( j \right) ) ( \overline{Y}_{2b}^{2b} )=0.  \\
\end{matrix}  \\
\end{matrix} \right.
\label{Eq11}
\end{equation}

This equation consists of a total of $4b^2$ equations and $n$ unknowns.
When $n>4b^2$, the equation is an underdetermined linear system. 
The SFG analytic weights $w_{j}^{(b)}$ represent the solution, which can be derived from this system of equations.

\begin{figure}[t]
    \centering
% 	\hrule height 5cm width 0.1pt
	\includegraphics[width=1\linewidth]{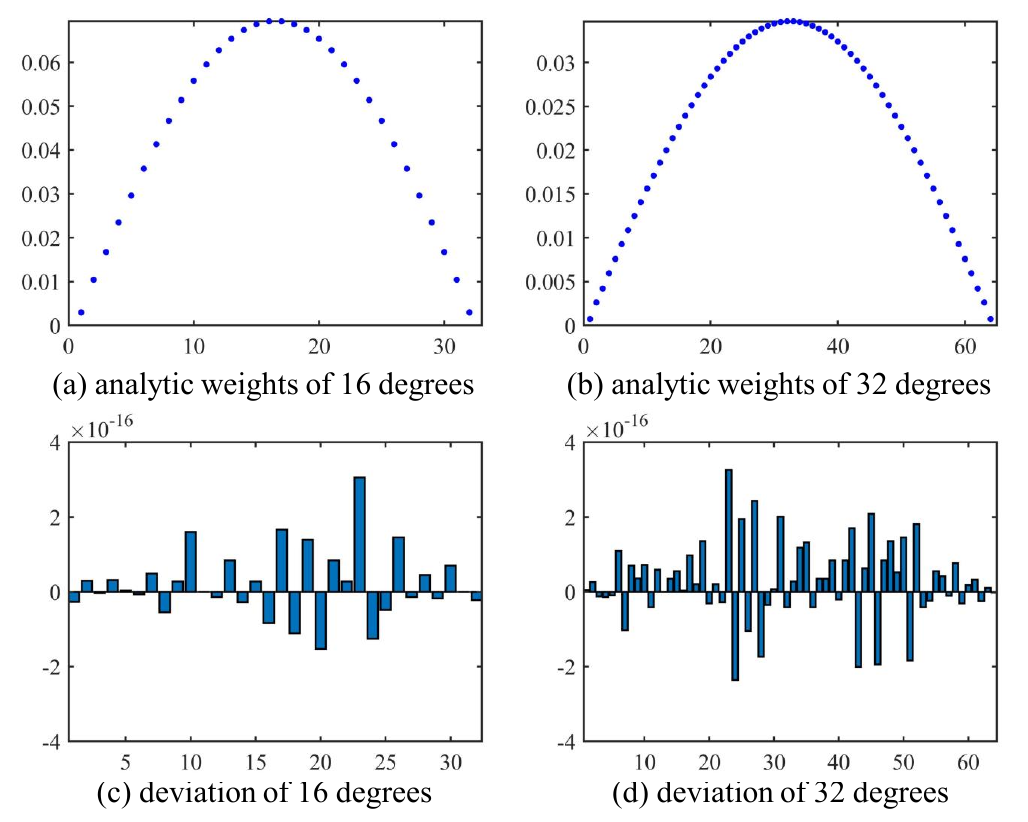}
	\caption{The values of analytic weights at 16 and 32 degrees, and the deviations between analytic weights and DH weights.}
\label{fig:analysis}
\end{figure}

Currently, the weights used in the SHT based on equiangular grids predominantly rely on \cite{driscoll1994computing}, referred to as DH weight. 
Since DH weight is effective in SHT, we verify the reliability of the analytic weight by comparing it with the DH weight. 
As shown in Fig.~\ref{fig:analysis}, the deviation between our analytic weight and the DH weight fluctuates on the order of $10^{-16}$. 
Thus, our method in this paper accurately determines the correct weight.

The SHT based on the SFG can be obtained. 
Let $f\left( \theta ,\phi  \right)$ be a band-limited function on $\mathbb{S}^2$, so when $l\ge b$, $\hat{f}\left( l,m \right)=0$,
\begin{equation}
\hat{f}\left( l,m \right)=\sqrt{2\pi }\underset{j=0}{\overset{n-1}{\mathop \sum }}\,w_{j}^{(b)}\overline{Y_{l}^{m}} ( {{F}_{\left[ \theta ,\phi  \right]}}( j ) ),
\label{Eq12}
\end{equation}
where $l<b,\left| m \right|\le l$, ${{F}_{\left[ \theta ,\phi  \right]}}$ is SFG function, and $w_{j}^{(b)}$ is the analytic weights obtained by the solution.

\begin{figure}[t]
    \centering
	\includegraphics[width=1\linewidth]{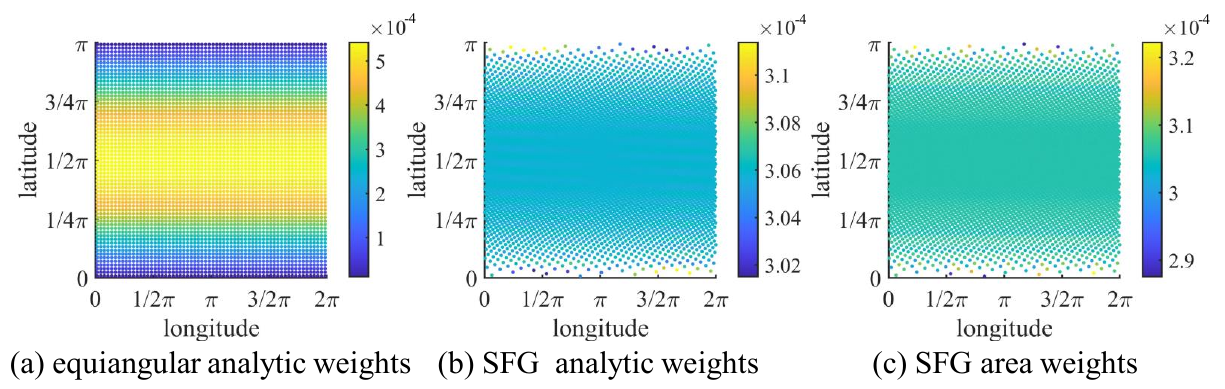}
	\caption{The distribution of the DH weights of equiangular (left), the analytic weights of SFG (middle) and  the area weights of SFG (right).}
\label{fig:DistributionWeights}
\end{figure}

Fig.~\ref{fig:DistributionWeights} illustrates the distribution of both the DH weights of equiangular, the analytic weights of SFG and the area weights of SFG.
Notably, the equiangular weights exhibit a peak near the equator and decrease towards the poles. 
Conversely, the weights of the SFG closely resemble equal weights, albeit with some fluctuations near the poles. 
These fluctuations serve to compensate for the discrepancies in the area enclosed by Fibonacci points and the higher-degree energy errors of spherical harmonics.

\section{Experiment and Analysis}
In this paper, we undertake the following evaluations and comparisons:
First, we assess the performance of analytic weights in the SHT, contrasting them with alternative weight schemes.
Next, we evaluate the ability of spherical harmonic representation under the SFG and compare it with other grid sampling methods.
Finally, we conduct two typical applications using the proposed FSH3D introduced in this study, comparing its performance to that of the traditional spherical harmonic approach.
These applications involve reconstructing shapes with star-shaped sets and classification using spherical harmonic descriptors.

\subsection{Accuracy Analysis of Analytic Weights}
\begin{figure}[t]
    \centering
	\includegraphics[width=1\linewidth]{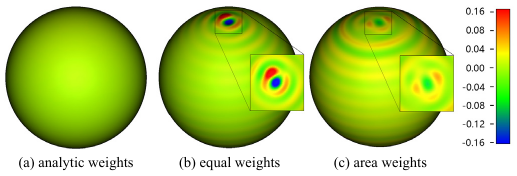}
	\caption{Error in reconstructing the unit sphere using analytic weights, equal weights, and area weights. Both equal weights and area weights exhibit deviations at the pole positions and introduce ripples. The SHT achieves optimal performance only when analytic weights are employed.}
\label{fig:SHrepresentation}
\end{figure}

The SHT is generally performed with either equal weights~\cite{swinbank2006fibonacci} or area weights~\cite{ahmad2007quasi}.
To validate the effectiveness of the analytic weights proposed in this article, we conduct forward and inverse SHT on the unit sphere, analyzing the error in reconstructing the unit sphere through SHT. 
Specifically, according to Shannon's sampling theorem, we sample 4300 points on the unit sphere using the SFG and compute the spherical harmonic coefficients from 0 to 31 degrees through SHT.
We utilize analytic weights, equal weights, and area weights to compute the spherical harmonic coefficients based on Eq.~(\ref{Eq12}). 
The analytic weights are obtained by solving each Fibonacci point according to Eq.~(\ref{Eq11}).
Equal weights assign identical weight to each point, while area weights correspond to the area of the Voronoi diagram where each Fibonacci point is situated. 
Fig.~\ref{fig:SHrepresentation} illustrates the error in reconstructing the unit sphere using analytic weights, equal weights, and area weights, with colors indicating the surface deviation.

\begin{figure}[t]
    \centering
	\includegraphics[width=1\linewidth]{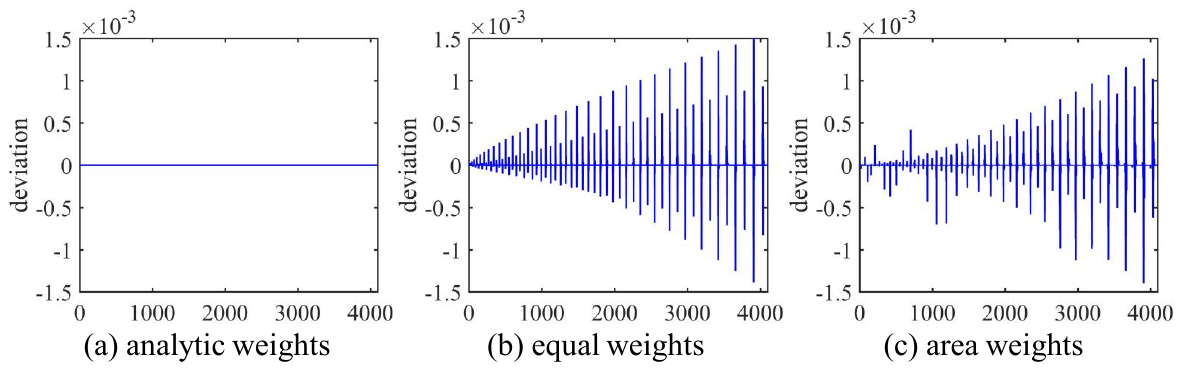}
	\caption{The deviation of ${{\hat{s}_{b}}}\left( l,m \right)$ about analytic weights, equal weights and area weights.}
\label{fig:deviation}
\end{figure}

In Fig.~\ref{fig:SHrepresentation}, both equal weights and area weights exhibit ripples in the error when reconstructing the unit sphere, particularly noticeable near the poles, indicating significant errors. 
While area weights demonstrate a partial improvement over equal weights in overcoming representation errors, they still fall short of achieving accurate representation. 
Optimal performance in SHT is attained only when analytic weights are employed.

According to the principles established in this paper, achieving a perfect expression of spherical harmonics necessitates satisfying the conditions outlined in Eq.~(\ref{Eq8}). 
Fig.~\ref{fig:deviation} displays the calculated deviation $\hat{s}_{b}'\left( l,m \right)-{{\hat{s}_{b}}}\left( l,m \right)$. 
The value of $\hat{s}_{b}'\left( l,m \right)$ is computed using analytic weights, equal weights, and area weights. 
It is observed that both equal weights and area weights fail to meet the conditions stipulated in Eq.~(\ref{Eq8}).
Consequently, their representation of spherical harmonics exhibits significant errors, indirectly affirming the pivotal importance of Equation Eq.~(\ref{Eq8}).

The choice of weights is crucial in computing spherical harmonic coefficients. 
Analytic weights offer superior compensation for deviations within the area enclosed by Fibonacci points and the higher-degree energy errors of spherical harmonics. 
Consequently, utilizing analytic weights significantly enhances the accuracy of reconstructing the unit sphere.

\subsection{Quantitative Analysis of SFG}
\begin{figure*}[!t]
    \centering
	\includegraphics[width=1\linewidth]{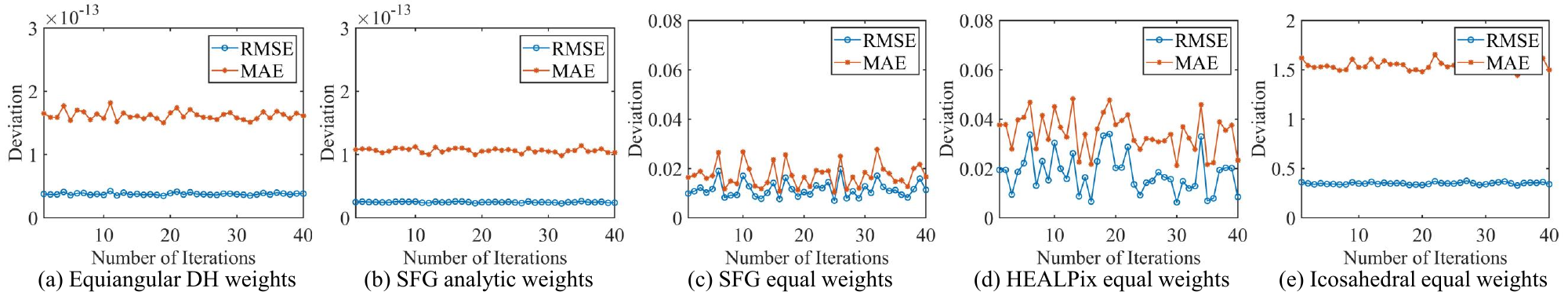}
	\caption{Quantitative comparison of equiangular grids, SFG, HEALPix grids, and Icosahedral grids. Note that the vertical scales of different charts are different, aiming to better display the details of each data.}
\label{fig:QuantitativeComparison}
\end{figure*}

A reliable SHT should exhibit minimal errors in spherical harmonic coefficients following both inverse and forward transform. 
With this in mind, we devised an experiment to verify the reliability of the FSH3D.
Initially, spherical harmonic coefficients are randomly generated. Then, these coefficients undergo an inverse transform followed by a forward transform, and the errors in the spherical harmonic coefficients before and after the transform are computed.

In the experiment, spherical harmonic coefficients are generated randomly 32 degrees.
We utilize equiangular grids, SFG, HEALPix grids, and Icosahedral grids. 
Equiangular grids partitioned a $64 \times 64$ sampling, generating 4096 points, while SFG employed 4096 points for sampling.
HEALPix and Icosahedral grids represent subdivision methods without the ability to specify the exact number of samples.
HEALPix grids are used with $S=20$, resulting in 4800 sampled points, while Icosahedral grids used $k=3$, leading to 5762 sampled points. 
Equiangular grids apply DH weights~\cite{driscoll1994computing}, SFG apply analytic weights, and other grids apply equal weights. 
We calculated root mean square error (RMSE) and mean absolute error (MAE) for spherical harmonic coefficients before and after different grid transform, conducting 40 sets of experiments for each.

In Fig.~\ref{fig:QuantitativeComparison}, the experiment evaluates spherical harmonic coefficients of 32 degrees across different grids. 
SFG, with analytic weights, exhibits the highest accuracy.
Compared to equiangular grids using DH weights, SFG reduces RMSE by approximately 34.6\% and MAE by approximately 34.0\%.
Uniformly distributed grids with equal weights address the uneven distribution of sampling points but fail to mitigate higher-degree energy errors of spherical harmonics. 
Compared to HEALPix, SFG exhibits similar but slightly superior quality. Although Fig.~\ref{fig:Distribution} shows that SFG and HEALPix have similar distributions, the SFG achieves a more uniform distribution over the entire sphere.
Consequently, grids with equal weights perform relatively poorly, yielding unstable results. 
The FSH3D method in this paper uses SFG with analytic weights. This approach employs a more uniform SFG, enhancing information comprehensiveness and overall stability in the SHT process.

\subsection{Reconstructing Shapes with Star-shaped Sets}
\begin{figure*}[!t]
    \centering
	\includegraphics[width=1\linewidth]{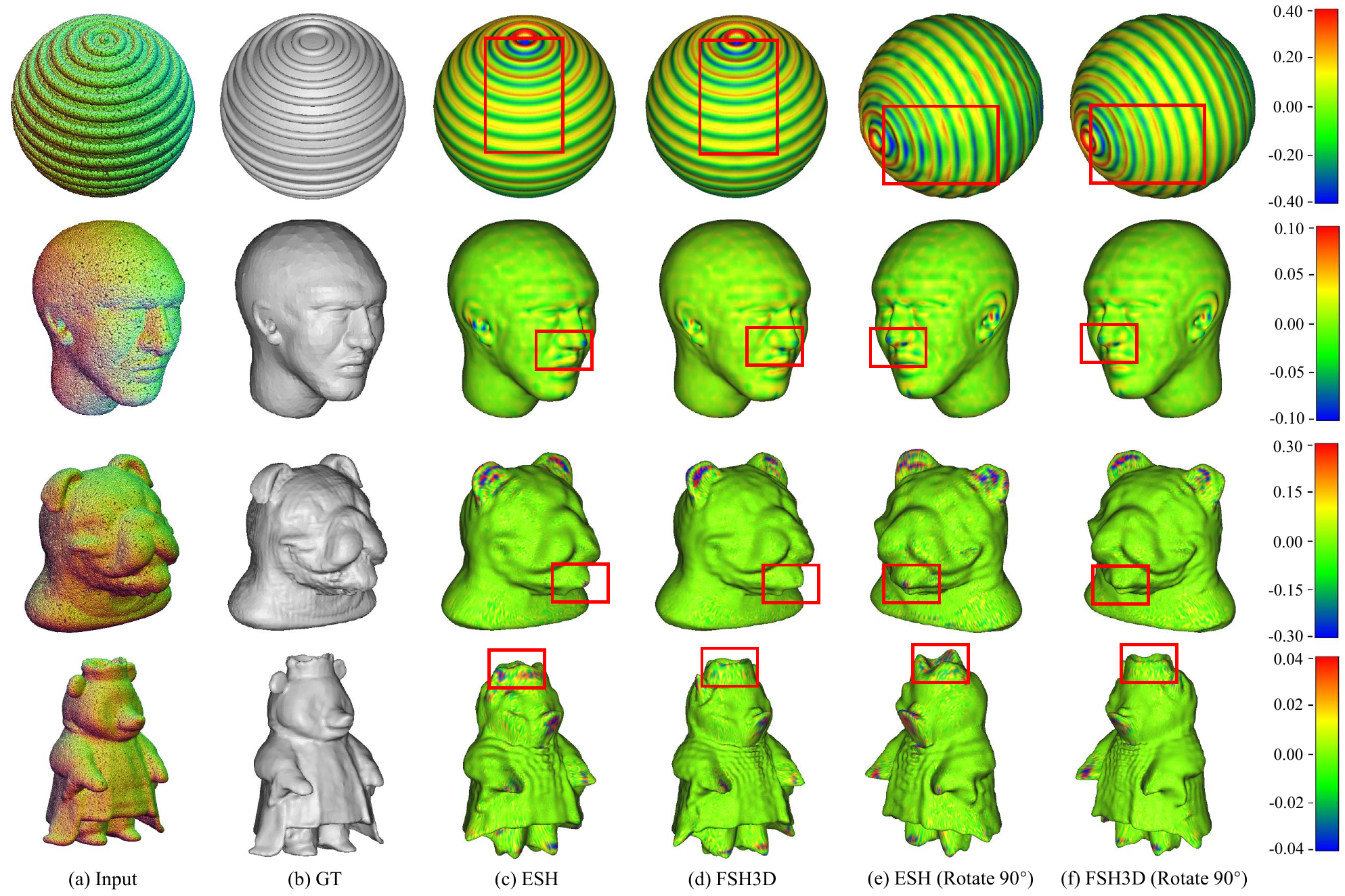}
    \caption{Visual comparisons of the reconstruction results of our method FSH3D (d)(f) and ESH (c)(e). (a) represents the input point cloud data, while (b) shows the corresponding mesh used as the ground truth. (c) and (d) depict the reconstruction results using (a) as the input data, whereas (e) and (f) present the reconstruction results when (a) is rotated by $90^\circ$ and used as the input. Deviations are measured using the Hausdorff distance, with the color map indicating reconstruction quality.} 
\label{fig9}
\end{figure*}

\begin{figure}[!t]
    \centering
	\includegraphics[width=0.8\linewidth]{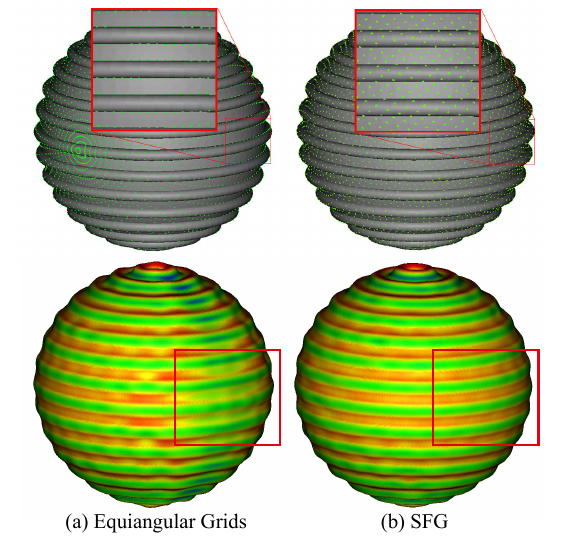}
	\caption{The sampling distribution of equiangular grids and SFG.}
\label{fig10}
\end{figure}

To assess the efficacy of FSH3D in 3D shape representation, we compile an experimental dataset. 
The dataset comprises 110 star-shaped shapes, manually created or sourced from the Princeton Shape Benchmark database, labeled as G1.
We normalize each shape in G1 according to the size of the bounding box, and the dataset G2 is generated by randomly rotating the shapes in G1. 
This is used to compare the effect of random rotation on the performance of FSH3D.
In the experiment, we sample both G1 and G2 datasets into point cloud data and represent them using spherical harmonics via SHT. 
For comparative analysis, we employ the currently optimal equiangular grids with DH weights (ESH) alongside SFG with analytic weights (FSH3D), with both methods calculating spherical harmonic coefficients up to 32 degrees.

We use the average deviations of RMSE, MAE, and volume error (VE) for two groups (G1 and G2) of shapes before and after rotation as evaluation indicators to assess the impact of rotation on FSH3D representation. 
The calculation method is as follows:
\begin{equation}
\begin{cases}
\text{RMSE}_{G1-G2} &= \sum_{i=1}^n\| \text{RMSE}_i^{G1}-\text{RMSE}_i^{G2}\|\\
\text{MAE}_{G1-G2} &= \sum_{i=1}^n \|\text{MAE}_i^{G1}-\text{MAE}_i^{G2}\|\\
\text{VE}_{G1-G2} &= \sum_{i=1}^n \|\text{VE}_i^{G1}-\text{VE}_i^{G2}\|
\end{cases}
\end{equation}
where, RMSE and MAE for each group of shapes are calculated by statistically analyzing the deviation in Hausdorff distance between the restored and original shapes. 
The VE is determined by calculating the volume deviation between the restored and original shapes.

Table~\ref{tab1} presents the deviations of RMSE, MAE, and VE before and after rotation. 
In most cases, the SFG exhibits smaller deviations in RMSE, MAE, and VE compared to equiangular grids. 
Thus, more uniform SFG sampling results in a more consistent spherical harmonic representation after rotation.
With the same number of sampling points, equiangular grids capture more features at the poles and fewer at the equator, providing advantages in certain scenarios.
However, after rotating the shape, the RMSE, MAE, and VE for equiangular grids change significantly. 
In contrast, the SFG maintains uniformity across the spherical surface, resulting in minimal changes in RMSE, MAE, and VE, and thus, a stable spherical harmonic representation regardless of rotation. 
Shapes with sharper features benefit more from SFG, showing consistent performance across different directions.
Equiangular grids sample more densely at the poles than at the equator. 
While this density can be advantageous for capturing key features at specific angles, uniform sampling of SFG provides more even feature capture and greater stability and versatility. 
Although the order of spherical harmonic coefficients limits shape detail, the uniform characteristics of SFG ensure that spherical harmonic representations at the same degree more closely resemble the original shape, preserving the shape more effectively.

\begin{table}[]
\setlength{\tabcolsep}{17pt}
\centering
\caption{Deviation of RMSE, MAE and VE before and after rotation of shapes. The units of RMSE, MAE and VE are $ 10^{-3}$.}
\label{tab1}

\begin{tabular}{l|cc}
\hline

                        & Equiangular grids & SFG \\ \hline
$\text{RMSE}_{G1-G2}$  & 1.083            & \textbf{0.087} \\
$\text{MAE}_{G1-G2}$   & 0.443             & \textbf{0.038} \\
$\text{VE}_{G1-G2}$    & 1.072             & \textbf{0.114} \\ \hline

\end{tabular}

\end{table}

Fig.~\ref{fig9} shows the visual results of  reconstructions of our method FSH3D, with all shapes are normalized and visualized from the same perspective. Our proposed method FSH3D excels in preserving details and remains unaffected by rotation. 
% As demonstrated in Fig.~\ref{fig9}, our proposed FSH3D method excels in preserving details and remains unaffected by rotation. 
There is a significant difference between Fig.~\ref{fig9}(c) and Fig.~\ref{fig9}(e). 
The ESH method employs uneven sampling, resulting in noticeable deformation when rotated, particularly evident as flattening at the equator in the first row. 
In contrast, Fig.~\ref{fig9}(d) and Fig.~\ref{fig9}(f) remain consistent under rotation due to FSH3D's more uniform sampling approach. 
This uniformity allows FSH3D to retain shapes better compared to equiangular grids.
As shown in Fig.~\ref{fig10}, equiangular grids exhibit noticeable deformation within the red-framed area, whereas SFG provides a more accurate representation in the same region.
The sparsity of the equatorial region in equiangular grids contrasts with the more uniform distribution of SFG, enabling it to capture richer details.

For finer representation, higher degrees of spherical harmonics are necessary, which require more sampling and computation. 
We investigate the representation effect of SFG at higher spherical harmonic degrees, using G1 as the experimental data.
In the experiment, spherical harmonic coefficients of 32, 64, and 128 degrees are used for the shapes in G1 to perform spherical harmonic representation using both equiangular grids and SFG. 
We compute the numerical deviation of the spherical harmonic representation from the original shape.

Table \ref{tab2} lists the RMSE, MAE, and VE for each experimental group. 
The FSH3D shows good performance across different degrees.
In spherical harmonics, low-frequency signals express overall features, while high-frequency signals capture more detailed features. 
Therefore, as the degrees of spherical harmonics increase, the deviation gradually decreases, resulting in a more refined representation. 
As the spherical harmonic degrees increase, SFG consistently outperforms equiangular grids, indicating that the method in this paper maintains good performance at higher spherical harmonic degrees.
Since spherical harmonic representation is a time-frequency domain analysis technique on a sphere, its capability increases with the degree. 
If the sampling is not uniform, representation for sharp features is insufficient. 
Conversely, the uniform SFG demonstrates less influence under rotation.

% 【表】 表2
\begin{table}[]
\setlength{\tabcolsep}{7.5pt}
\centering
\caption{Evaluation metrics from 110 star set shapes, including RMSE, MAE and VE. The units of RMSE, MAE and VE are $ 10^{-3}$.}
\label{tab2}
\begin{tabular}{clS[table-format=2.3]S[table-format=2.3]S[table-format=2.2]}
\hline
Degrees                     & Methods           & {RMSE} & {MAE} & {VE} \\ \hline
\multirow{2}{*}{32}         & Equiangular grids & 14.902           & 10.612           & 75.12           \\
                            & SFG   & 14.577           & 8.901          & 3.80            \\ \hline
\multirow{2}{*}{64}         & Equiangular grids & 5.598            & 1.754           & 19.50           \\
                            & SFG   & 4.886            & 1.457           & 3.85            \\ \hline
\multirow{2}{*}{128}        & Equiangular grids & 2.702            & 0.665           & 2.23            \\
                            & SFG   & 1.467            & 0.353           & 0.58            \\ \hline
\end{tabular}
\end{table}

\subsection{Classification through Spherical Harmonic Descriptors}
Spherical harmonics-based classification and retrieval methods typically use SHT to obtain spherical harmonic coefficients, which are then used to construct descriptors.
Two distinct approaches are commonly employed for this purpose:
The conformal parameterizations spherical harmonic descriptor (CPSHD) method, as described in \cite{agus2020wish}, involves achieving a spherical parameterization of the input mesh through Willmore flow. This parameterization is then utilized to create the spherical harmonic descriptor.
The spherical harmonic descriptor (SHD) method, introduced in~\cite{funkhouser2003search}, decomposes the 3D shape into a series of shells, from each of which spherical harmonic features are extracted.

\begin{figure}[!t]
    \centering
	\includegraphics[width=1\linewidth]{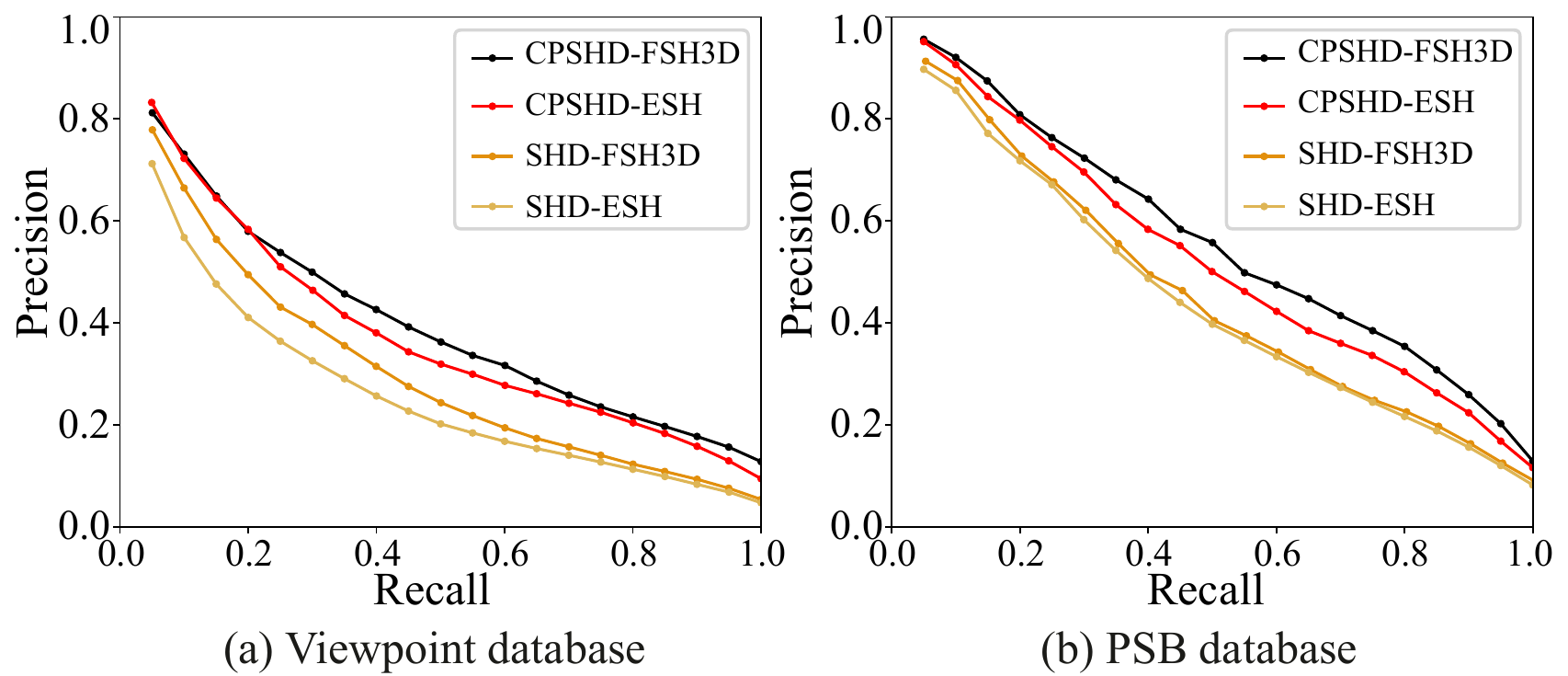}
	\caption{Precision-Recall curve performance comparison of spherical harmonic shape descriptors.}
\label{fig_kj}
\end{figure}

In order to better showcase our performance, we choose two types of databases, that is, the Viewpoint database and the Princeton Shape Benchmark(PSB) database~\cite{shilane2004princeton}.
These two databases offer multiple semantic labels for each shape. 
For a given shape to be detected, these comparative methods calculate descriptors to subsequently ascertain the similarity between the 3D shape to be detected and the each labeled shape.
In the experiment, we used both the ESH and FSH3D methods to construct four spherical harmonic descriptors: SHD-ESH, SHD-FSH3D, CPSHD-ESH, and CPSHD-FSH3D. 
We then conducted classification experiments using these four descriptors. 
Fig.~\ref{fig_kj} presents the experimental results using a general Precision-Recall curve for 3D shape classification, based on all objects in both databases. 
Generally, a curve closer to the outside indicates better performance.

The results indicate that SHD-FSH3D outperforms SHD-ESH, and similarly, CPSHD-FSH3D performs better than CPSHD-ESH. This demonstrates that the FSH3D method enhances the performance of both descriptors.
Further analysis by category shows that the improvement is not particularly significant for objects such as balls, blocks, and beds. 
However, for objects with more prominent sharp features, such as chairs and knives, the improvement is more pronounced. 
This is because FSH3D sampling is more uniform, allowing for better preservation and more stable representation of sharp features, consistent with our previous descriptions.

\section{Conclusion}
In this work, we propose a 3D representation method based on Fibonacci Spherical Harmonics (FSH3D), which achieve more accurate 3D representation through a uniform SFG sampling technique. 
Additionally, we evaluate FSH3D in two applications: 3D shape reconstruction and 3D shape classification. 
The results demonstrate that FSH3D offers more accurate 3D representation and superior rotation invariance.
Spherical harmonics, as an implicit representation method, are highly suitable for shape reasoning applications in learning frameworks. 
FSH3D effectively enhances the stability of frequency domain representation under rotation transformation. 
In future work, we will explore incorporating spherical harmonics into learning frameworks to further improve 3D representation capabilities.

\section*{Acknowledgments}
This work was supported by the National Natural Science Foundation of China (No. 92367301, No. 92267201, No. 52275493, No. 92160301).

%-------------------------------------------------------------------------
% bibtex
\bibliographystyle{eg-alpha-doi} 
\bibliography{egbibsample}       
%-------------------------------------------------------------------------

\end{document}